\documentclass[12pt, final, oneside]{article}

\usepackage{cite}
\usepackage[english]{babel}
\usepackage[affil-it]{authblk}
\usepackage{times}
\usepackage{amsmath}
\usepackage{amsfonts}
\usepackage{amssymb}
\usepackage{svg}
\usepackage{subfig}
\usepackage{enumitem}
\usepackage[hidelinks]{hyperref}
\usepackage{xr}

\makeatletter

\renewcommand{\p@figure}{fig. }
\renewcommand{\p@table}{tbl. }
\renewcommand{\p@section}{sec. }
\renewcommand{\p@subsection}{sec. }

\makeatother

\title{Effects of the entropy source on Monte Carlo simulations}

\date{}

\begin{document}

\author[1]{Anton Lebedev}
\author[2]{Annika M\"{o}slein}
\author[1]{Olha I. Yaman}
\author[3]{Del Rajan}
\author[3]{Philip Intallura}

\affil[1]{The Hartree Centre, Keckwick Ln, Warrington, UK\footnote{\{anton.lebedev, olha.ivanyshyn-yaman\}@stfc.ac.uk}  } 
\affil[2]{Quantum Dice Limited, 264 Banbury Road, Oxford,  UK\footnote{annika.moeslein@quantum-dice.com} }
\affil[3]{HSBC Lab, Innovation \& Ventures, 8 Canada Square, London, UK\footnote{del.rajan@hsbc.com} } 

\maketitle

\begin{abstract}
In this paper we show how different sources of random numbers influence the outcomes of Monte Carlo simulations.
We compare industry-standard pseudo-random number generators (PRNGs) to a quantum random number generator (QRNG) and show, using examples of Monte Carlo simulations with exact solutions, that the QRNG yields statistically significantly better approximations than the PRNGs. Our results demonstrate that higher accuracy can be achieved in the commonly known Monte Carlo method for approximating $\pi$.
For Buffon's needle experiment, we further quantify a potential reduction in approximation errors by up to $1.89\times$ for optimal parameter choices when using a QRNG and a reduction of the sample size by $\sim 8\times$ for sub-optimal parameter choices. We attribute the observed higher accuracy to the underlying differences in the random sampling, where a uniformity analysis reveals a tendency of the QRNG to sample the solution space more homogeneously.
\end{abstract}

\tableofcontents

\section{Introduction}
Stochastic simulations have been one of the earliest applications of large-scale computer simulations in today's science and industry. Monte Carlo (MC) simulations, which use random numbers as an integral part of obtaining a solution of stochastic problems, were first developed by Neumann and Ulam during the Manhattan project, and are used to this day in planning radiation therapy \cite{nationalresearchcouncilofcanada.metrologyresearchcentre.ionizingradiationstandardsEGSnrcLogicielPour2021},
protein folding \cite{sanghyunExtractingEquilibriumNonequilibrium2004}, particle methods in general \cite{weigelUnderstandingPopulationAnnealing2021} and the modelling of societal segregation \cite{ParablePolygons2, clarkResidentialPreferencesNeighborhood1991}. Evolution of stochastic processes, described by stochastic differential equations (SDE), is a mainstay of the financial industry (such as the Black-Scholes equation), and underpins the development of optical systems 
\cite{mullerLaminarChaos2018, sorianoComplexPhotonicsDynamics2013} as well as high-precision measurements \cite{braunHeisenberglimitedSensitivityDecoherenceenhanced2011}. The numerical solution of a stochastic differential equation is a specific case of a Monte Carlo method.
Additionally, the growing adoption of uncertainty quantification, to include uncertainties in modelling as well as the stochastic nature of real-world observations, introduces Monte Carlo methods to areas in which these have previously played only a niche role.

Fundamental to all of the fields listed above is the availability of random numbers on the platform used for the simulations.
It is, however, commonly known that all "random" numbers available in system libraries are simulated by algorithms generating deterministic sequences and are therefore not truly random, \cite{Lecuyer}. Mistakes in the past, such as highly correlated RANDU generator\cite{Randu}, have resulted in the development of high-quality pseudo-random number generators (PRNGs), of which the Mersenne Twister (MT) \cite{MersenneT} is nowadays seen as the industry standard, being the standard engine adopted in Python \cite{RandomGeneratePseudorandom} (and NumPy up to v. 1.15 \cite{RandomGeneratorNumPy}) and a common example introducing PRNGs in C++\cite{meyersEffektivProgrammieren552011}. 
Scientific applications of Monte Carlo methods, relying on repeating the same operations,  are especially popular due to the relative ease of their implementation and parallelisation, allowing for a simple utilisation of all available compute resources. However, the PRNGs described above are, in general,
not suited for practical parallel applications. To fill this void, parallel PRNG libraries exist, 
such as Tina's random number generator library or the scalable parallel random number generator library \cite{bauke2011tina}.

Given some parameters, the PRNG will yield a number from a periodic sequence of numbers. Generally, the
parameters are a \textit{seed} and the iteration number, although the latter is often kept as an internal counter of the RNG. The seed determines the starting position within the sequence, where a fixed seed will always result in the same sequence of numbers. Usage of a fixed seed facilitates testing of fundamental software functionality and thus software development, but is highly questionable for the purposes of statistics.

Progress in quantum technology over the last decade has led to the commercial availability of quantum random
number generators (QRNGs), whose reliance on fundamental principles of indeterminacy in quantum physics
can yield high-quality random numbers. Especially self-certifying optical QRNGs promise the genuine extraction of randomness from quantum systems isolated from electronic noise, while further operating at high throughput \cite{PhysRevX.10.041048}. 

Although a variety of random number sources exists nowadays, the quality measurements of their outputs are
surprisingly limited. The Dieharder suite is a commonly used test suite to verify whether
a RNG provides sufficiently random bit sequences \cite{Diehard, Dieharder}. The usefulness of this test suite in the context of Monte Carlo simulations is debatable and
attempts have been made in the past
to investigate potential performance advantages of QRNGs with, e.g., the Ising model \cite{IsingRNG, Ghersi2017}. Results of these investigations are generally limited
to a qualitative statement and, to the authors' best knowledge, do not include proper validation. 

Even though reference examples, where benefits of QRNGs in MC simulations have been demonstrated, exist\cite{CirauquiQRNGPRNG2024, Ghersi2017}, one might argue that benchmarking against PRNGs exhibiting periods shorter than the period of the ’minimum standard’ adopted in C++ \cite{ParkMillerMinStandard} overemphasises the benefits of QRNGs.
In principle, as has been demonstrated by Cirauqui et al. in \cite{CirauquiQRNGPRNG2024}, problems sensitive to correlations are suitable for an analysis of RNG quality. In the case delineated therein, the observables are computed from spins which, in turn, are directly dependent on the sequence of random numbers provided and thus an effective proxy measure for correlations within the RN sequence.

In this paper, we show that a QRNG will yield a better approximation than an industry-standard PRNG for a much simpler and cheaper (pertaining to the run time) model.

In the following, we introduce models used to investigate the suitability of various random number sources for Monte Carlo applications as well as the tests used to ascertain the significance of the observations.

\section{Results}
\begin{figure}
    \centering
    \includegraphics[width=0.9\linewidth]{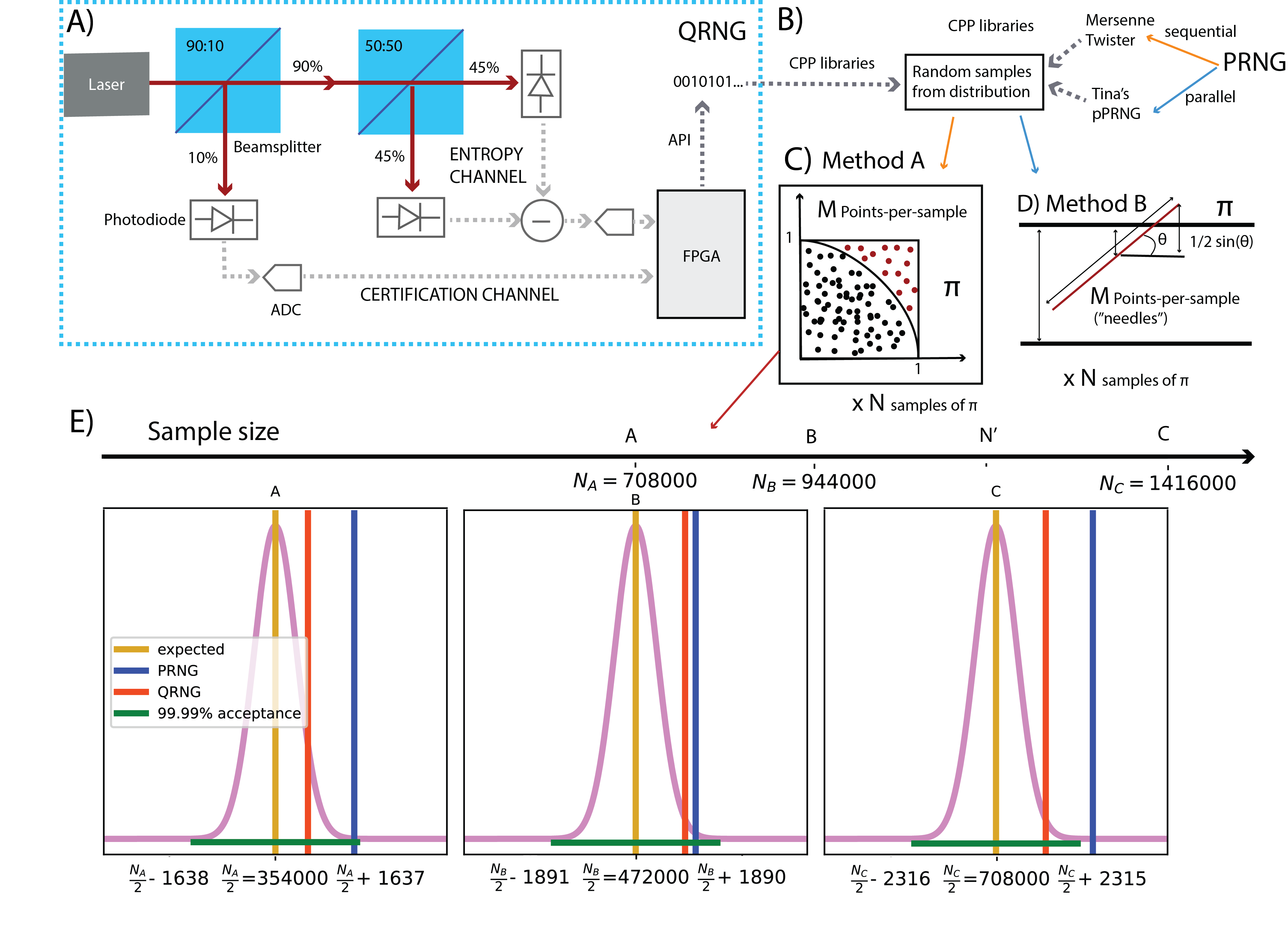}
\caption{A) Setup of the employed QRNG leveraging the inherently unpredictable paths of photons: measuring the differencing signal between two beamsplitter outputs and postprocessing generates a bitstream of random numbers, self-certified in real-time\cite{PhysRevX.10.041048}. B) The stream of random bits from the QRNG or PRNG is transformed into a list of values conforming to the distributions of random numbers using the C++ standard library \cite{CPPLibr}. C) Monte Carlo method to approximate $\tfrac{\pi}{4}$ (Method A). D) Monte Carlo method to approximate $\tfrac{2}{\pi}$ (Method B, "Buffon's needle". E) The evolution of the sign test for growing sample size $N$ of data obtained using Method A.}
    \label{fig:EvolutionOfTheBinomialTest}
\end{figure}

A MC simulation samples, in general, an unknown distribution thereby providing an empirical
approximation to said distribution. We use the archetypal MC \textit{approximation of} $\pi$ as the initial stochastic test bed for our analysis to assess how different sources of randomness affect the simulation outcomes.
This choice is motivated by the availability of an exact solution, simplicity of the procedure as well as the model being the universal example for MC methods. We employ the perhaps best known $\tfrac{\pi}{4}$ approximation (Method A, c.f. \ref{fig:EvolutionOfTheBinomialTest}C), i.e. generating random points inside a unit square and counting the proportion that falls inside the quarter circle, as well as "Buffon's needle" approximation
of $\tfrac{2}{\pi}$ (Method B, c.f \ref{fig:EvolutionOfTheBinomialTest}D)\cite{Buffon}. Due to the different quantities being approximated ($\tfrac{\pi}{4}\;\text{vs.}\;\tfrac{2}{\pi}$), we expect the uncertainties of the resulting $\pi$ approximations to behave differently with respect to the number of points $M$ on $[0,1]^2$ used to obtain one sample $\tilde{\pi}_i$ of $\pi$ as well as the number
of samples $N$ used to obtain an approximation of $\bar{\pi} = \tfrac{1}{N}\sum_{i=1}^N \tilde{\pi}_i$. 
Comparing a self-certifying quantum random number generator (c.f. \ref{fig:EvolutionOfTheBinomialTest}A and Supplementary Materials SM sec. \ref{QRNGmethod}) to industry-standard PRNGs (SM sec. \ref{PRNGmethod}), we demonstrate that the QRNG leads to better approximations than the PRNGs for both methods. We show that the results obtained with the QRNG pass the sign test and t-tests for larger sample sizes than the PRNG: on the one hand, this permits a reduction in sample size while maintaining the same accuracy as achieved by the PRNG. On the other hand, a better approximation of $\pi$ is achieved when considering the same number of samples. 

To investigate the observed differences further, we then perform an uniformity analysis to assess underlying sampling itself.

\paragraph{On Reproducibility}\label{subsec:Reproducibility}
Fundamentally, when using a true random number generator (classical or quantum) the simulation results are not
bit-wise reproducible, i.e., two consecutive executions will yield slightly different outcomes. An outcome of a stochastic simulation is deemed reproducible if - all parameters being equal - the results of repeated execution yield distributions of the desired observables that do not differ in a statistically significant way. To ensure reproducibility of our data collection and analysis procedures we utilise the Snakemake\cite{molderSustainableDataAnalysis2021} framework to automate the compilation of the collection routines and the actual data collection.
\paragraph{Monte Carlo estimates of $\mathbf{\pi/4}$}

First we consider the commonly known method of estimating $\pi$ (Method A, c.f. SM \ref{subsect:pi_approximation}) and apply the sign test to $N=2.832\cdot 10^{6}$ $\tilde{\pi}$ samples (c.f. SM \ref{subsec:Tests}), each obtained using $M=10^5$ points on $[0,1]^2$. The coordinates $(x,y)$ of each point are drawn from the uniform distribution between 0 and 1 using a QRNG or Mersenne Twister PRNG, respectively. We consider the $99.99\%$ confidence interval of the hypothesis $H_0: m=\pi$ (i.e. the median is $\pi$), with the alternative hypothesis $H_A:m\neq \pi$. Here, \ref{fig:EvolutionOfTheBinomialTest} E illustrates the evolution of the acceptance interval (shown as the green bar) for growing sample sizes, along with the values of the test statistic (number of values $>m$) for the datasets obtained with PRNG (blue) and QRNG (red) and the ideal statistic ($\frac{N}{2}$, gold).
The following can be observed: after roughly $N_C \sim 1.4  \cdot 10^6$ $\tilde{\pi}$ samples, $H_0$ must be \textit{rejected} for the data obtained using the Mersenne Twister PRNG (i.e. the test statistic leaves the acceptance range) but it can \textit{not} be rejected for data obtained using the QRNG. %
Explicitly this means that we must reject the claim that the median is $\pi$ for PRNG data but do not reject it for QRNG data: the latter may thus be considered to provide qualitatively better results

We emphasise that no statement is made about a statistical error of the 2nd kind ($H_A$ is true but $H_0$ is accepted) for the QRNG dataset, i.e., the analysis does not bound the probability of $H_A:m\neq\pi$ for the QRNG dataset.

With $N_C$ corresponding to the entire obtained dataset, we determined that the hypothesis is consistently rejected for PRNG data starting with $N'\approx 0.82 N_C$ samples, whereas QRNG data remains accepted. This implies that for $N>N'$  only the QRNG data yields acceptable results. We make no statement on the quality of each approximation for $N\leq N'$, where both datasets are accepted. This observation means that if the sample sizes are increased beyond $N'$ (tested for $2\times N_C$), higher accuracy of the simulation results can be achieved with QRNG input, whereas no improvement in accuracy can be obtained with the PRNG. 

\paragraph{Buffon's needle}

\begin{figure}[htb]
    \centering
    \includesvg[width=0.6\linewidth]{picts/PI_DSMCvsBUFFON_StandardisedError_PPSdependence.svg}
    \caption{Dependence of the approximation error $\tilde{\epsilon}$ on points per $\tilde{\pi}$ sample.}
    \label{fig:StdErrorPPSDependence_BuffonVsSimple}
\end{figure}

Since Buffon's method (method B) approximates $\frac{2}{\pi}$ instead of $\frac{\pi}{4}$, the behaviour of the $\tilde{\pi}$ distributions is different. In the Supplementary Materials \ref{fig:SimpleVsBuffonDistribution}, we show that the mean obtained with method B differs from the true value of $\pi$ more than the mean obtained with method A, indicating that method A is the better choice for approximating $\pi$ via MC simulations. However, although method B yields larger approximation errors for $M\leq 5\cdot10^5$, it displays a clear dependence on the number of points-per-sample (pps) $M$ by following the expected $\frac{1}{\sqrt{M}}$ convergence, resulting in a better approximation at $M=10^6$ pps and beyond (c.f. \ref{fig:StdErrorPPSDependence_BuffonVsSimple}). In conjunction with the analysis presented below, we conclude that method B allows for a clear distinction of the effects of the sample size $N$ and the points-per-sample $M$, whereas for method A such a distinction is not easily made.
\begin{figure}
    \centering
    \subfloat[$M=10^3$]{
        \includesvg[width=0.5\linewidth]{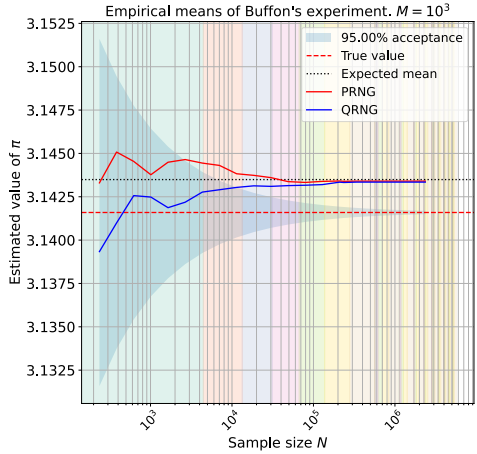}
        \label{subfig:EmpiricMeansEvolution_M1k}
    }
    \subfloat[$M=10^5$]{
        \includesvg[width=0.5\linewidth]{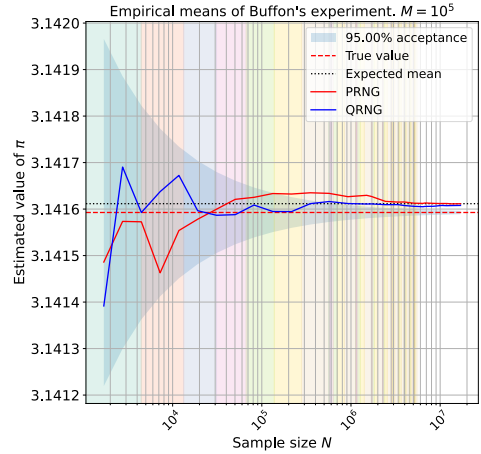}
        \label{subfig:EmpiricMeansEvolution_M100k}
    }
    \\
    \subfloat[Deviation (error) for $M=10^5$]{
    \includesvg[width=\linewidth]{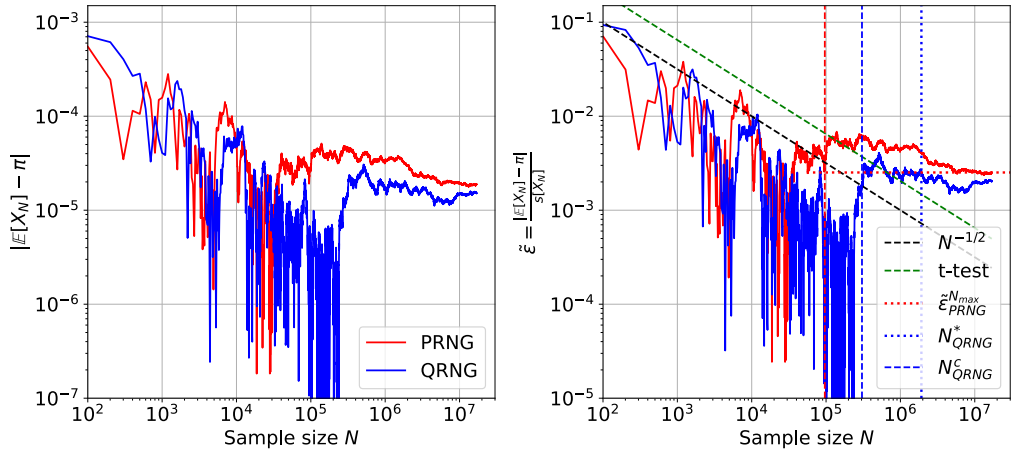}
    \label{subfig:DeviationFromPi_100k}
    }
    \caption{a,b) Evolution of the empirical means for QRNG and PRNG data with growing sample size $N$ at a fixed pps $M$. Note that the ranges of the y-axis are different for $M=10^5$ and $M=10^3$. c) Deviation (error) of the computed approximation from $\pi$ using $M=10^5$ pps (left: absolute error to exhibit differences without any scaling; right: normalised by the standard deviation for comparison to convergence metrics.)}
    \label{fig:EmpiricMeansEvolution}
\end{figure}

In \ref{fig:EmpiricMeansEvolution} the behaviour of the empirical mean $\mu$ with growing sample size for a dataset of $N=1.6\cdot 10^7$ samples using $M=10^5$ pps, and $N=2.5\cdot 10^6$ samples using $M=10^3$ pps is shown for both QRNG and parallel PRNG (pPRNG).

The funnel-shaped shaded blue region corresponds to the $95\%$ acceptance range obtained by reordering the $t$-test equation (c.f. SM eq. \ref{eq:T_test_statistic}) with the upper and lower bounds obtained utilising the empirical variance of the joint pPRNG and QRNG dataset. Additionally, multi-coloured regions of the x-axis indicate within which execution of the code the data has been obtained, each of which corresponds to a pPRNG initialisation with a different seed (c.f. SM \ref{sec:numerics}).
We observe in \ref{fig:EmpiricMeansEvolution}b that the pPRNG mean is consistently rejected as an approximation of $\pi$ after $N=2\cdot 10^5$ samples, whereas the QRNG mean is rejected only after $N=10^6$ samples. Here we must note that the values converge to the true value only in  the case $M\rightarrow \infty$, while, for a finite $M$, the limit value is unknown but larger than $\pi$. We have devised an estimator (c.f. SM eq. \ref{eq:estimate_2}) for the difference between the limit value and $\pi$ at a given pps $M$ in SM \ref{subsect:pi_approximation}. The expected mean is shown as the black dotted line in \ref{fig:StdErrorPPSDependence_BuffonVsSimple} and is in good agreement ($\mu_{data}-\mu_{expect} = 3.8\cdot 10^{-6}$)  with the empirical mean taken across both datasets.\\

In spite of the unknown limit value we use the method to assess the difference between PRNG and QRNG, as the values converge to $\pi$ in the limit $N \cdot M\rightarrow \infty$ (c.f SM \ref{fig:Convergencebehaviour}), and a comparison of the random number sources for identical parameters remains appropriate.
Computing the normalised deviation $\tilde{\epsilon}_N:=\frac{|\mathbb{E}[X_N] - \pi|}{\sigma[X_N]}$ yields the results exhibited in \ref{fig:EmpiricMeansEvolution}c.

Here, evidently, the approximation error (normalised deviation) when using Buffon's method decreases with growing number of samples $N$, as predicted by the central limit theorem. However, after $N_{PRNG}^C\approx 10^5$ samples obtained with the pPRNG ($N_{QRNG}^C\approx3\cdot 10^5$ for QRNG), the behaviour changes drastically, converging to a limit value $\tilde{\epsilon}\approx 2\cdot 10^{-3}$ (c.f. \ref{fig:StdErrorPPSDependence_BuffonVsSimple}). 

We denote the largest sample size after which $\tilde{\epsilon}_N$ exceeds the $1/\sqrt{N}$ threshold as $N^C_{QRNG,PRNG}$. Beyond this threshold $\tilde{\epsilon}_N$ breaks the expected convergence behaviour and t-test acceptance bounds (indicated by the dashed lines in the plot), thus resulting in no further increase in accuracy.
We note that $N^C_{PRNG}\approx M$ for various $M$ and, in view of \ref{fig:StdErrorPPSDependence_BuffonVsSimple}, interpret that the points-per-sample $M$ dictate the "trueness" of the approximation, whereas the number of samples $N$ determines the precision of the measurement (c.f SM \ref{fig:Convergencebehaviour}). For $N\leq M$ the accuracy is dictated by the trueness of the observation, whereas for $N>M$ it is determined by the precision of each observation.
This implies that the number of points-per-sample $M$ must scale as $\sim N$ to maintain the $\sim 1/\sqrt{N}$ convergence. Hence, a simulation with $N > N^C\approx M$ is of little use as the approximation error plateaus for $N>M$ (c.f. \ref{fig:EmpiricMeansEvolution}c).
The same holds true for method A, albeit there the distinction between $M$ and $N$ is less clear as both affect the width of the distribution.

Use of the QRNG maintains the expected convergence behaviour up to $N_{QRNG}^C = 304201$ and thus $\approx 3.16\times$ longer than the pPRNG ($N^C_{PRNG}=96201$). This leads to higher accuracy of the samples obtained with the QRNG, with the resulting approximation error $\tilde{\epsilon}_{N_{QRNG}^C}$ being $\approx 2.99\times$ smaller than $\tilde{\epsilon}_{N_{PRNG}^C}$. 

As shown in SM \ref{fig:SimpleVsBuffonDistribution}, the samples are approximately normally distributed and as such use of the t-test is appropriate. To obtain the $95\%$ acceptance threshold of the t-test, the $\frac{1}{\sqrt{N}}$ curve must be scaled by $2.05$ (c.f. SM eq. \ref{eq:T_test_statistic}) resulting in the green dashed line in \ref{fig:EmpiricMeansEvolution}c. The range of acceptable deviations is thus larger: when we apply the same analysis as above, this results in $N_{PRNG}^C=2.3\cdot 10^5 = 2.3 M$  and $N_{QRNG}^C=10^6 = 10M$. Thus using the QRNG results in a $\sim 4.54\times$ larger admissible sample size and an approximation error which is $\sim 1.89\times$ smaller than that of the pPRNG approximation.

In practice, the asymptotic behaviour $\sim 1/\sqrt{N}$ is used to estimate the sample size for which the error will fall below the desired threshold.  Since the  exact dependence of the approximation error on $N$ (and more so on $M$) is generally unknown, the sample size $N_{max}$ is generally generously over-estimated. Thus, in practice, one is likely to use a sub-optimal $N_{max} \gg N^C$.
Given such a choice, the approximation error $\tilde{\epsilon}_{PRNG}^{N_{max} }$ of the entire pPRNG dataset of $N_{max}=1.69\cdot 10^7$ is illustrated by the horizontal red dotted line in  \ref{fig:EmpiricMeansEvolution}c. Here, we observe that the same accuracy is already attained at $N_{QRNG}^{\ast}\approx 1.92\cdot 10^{6}$ samples using the QRNG (vertical dotted line), or in other words, the same accuracy is already reached using the QRNG with a $\sim 8.3\times$ smaller sample size. Keeping $N_{max}$ fixed, the approximation error of the QRNG data computed is $\tilde{\epsilon}_{N_{max}} / \tilde{\epsilon}_{N_{QRNG}^{\ast}} = 1.25$ times smaller than that of the pPRNG data.
%


\begin{figure}[htb]
    \centering
    \includesvg[width=0.6\linewidth]{picts/DSMCPI_StandardizedDeviationFromPi_3RNGs100000_Final.svg}
\caption{Deviation (error) of the computed approximation from $\pi$ at a fixed pps $M$ for data obtained with the QRNG and two different PRNGs.}
    \label{fig:3RNGsCompared}
\end{figure}

In \ref{fig:3RNGsCompared}, we augment fig. \ref{subfig:DeviationFromPi_100k} to contrast the behaviour of the QRNG (blue) and pPRNG (red), exhibiting a large period, to that of the "Minimum standard" linear congruential generator\cite{ParkMillerMinStandard} (LCG, black) as it is implemented in the C++11 standard. First note that the LCG dataset size was capped at $1/3$ of the pPRNG dataset size since the approximation error does not decrease beyond $\sim N_{QRNG}^C$. The normalised deviation $\tilde{\epsilon}_N$ displays several notable features for the LCG that are not present in either the QRNG or pPRNG equivalents.
For $N>N_{QRNG}^C$, it reaches, on average, an asymptotic value of $\tilde{\epsilon}_{LCG} = 1.24\cdot 10^{-2}$, or $\approx 6.12\times$ larger than the deviation obtained with QRNG data. 

Further we observe that $N_{LCG}^C\approx 1.9\cdot 10^{4}$  which is $\sim 12\times$ smaller than $N_{PRNG}^C$. Finally, $\tilde{\epsilon}_{LCG}$ displays a periodic structure for $N\gtrsim N_{LCG}^C$, whose oscillation magnitude decreases, but does not vanish, with increasing sample size. The oscillation period is directly related to the period $m=2^{31}-1\approxeq 2.15\cdot 10^{9}$ of the LCG, which, for the given $M$, translates to a period $N\approxeq 10737$ of the $\tilde{\pi}$. Given the notably higher approximation error $\tilde{\epsilon}$, $N_{LCG}^C$ being an order of magnitude lower than $N_{PRNG}^C$, and the obvious effects of the LCG period, we have excluded this LCG from our experiments and, in contrast to other works \cite{CirauquiQRNGPRNG2024}, not considered PRNGs of shorter periods than the "Minimum standard".

\paragraph{Uniformity analysis}
To better understand the observations made with the $\pi$ models, we consider the underlying sampling, or more specifically, the uniformity of
the point distribution on $[0,1]^2$ as provided by the QRNG and the Mersenne Twister\cite{MersenneT} PRNG. 
Uniformity was analysed using the nearest-neighbour distance ($N\!N$), largest empty sphere ($LES$), dispersion ($C_V$, $C\!E$) and potential
energy metrics ($Q$ and $R_{nr}$)(SM \ref{subsec:uniformitymethod}). We remark that an analysis of the uniformity of sampling of a domain requires an analysis of the relation of the point distribution to the boundary. In absence of such the homogeneity of the distribution is considered. Hence, in the following we refer to solely the behaviour of a point distribution, where $N\!N$ and $LES$ assess the distance between pairs of points, $C\!E$ and $R_{nr}$ test for clustering and 'non-randomness', and $C_V$ and $Q$ provide more global measures for uniformly dispersion by averaging the contributions from each point. These measures, further described in SM \ref{subsec:uniformitymethod}, are collected 50 times for the sample points sets of various sizes and numerical precision both for sequential pre-generated QRNG and PRNG (Mersenne Twister). The 50 averages, each taken from 100 experiments with a given number of point samples, are summarised in the box plots.

\begin{figure}[ht!]
    \centering
    \includegraphics[width=1\linewidth]{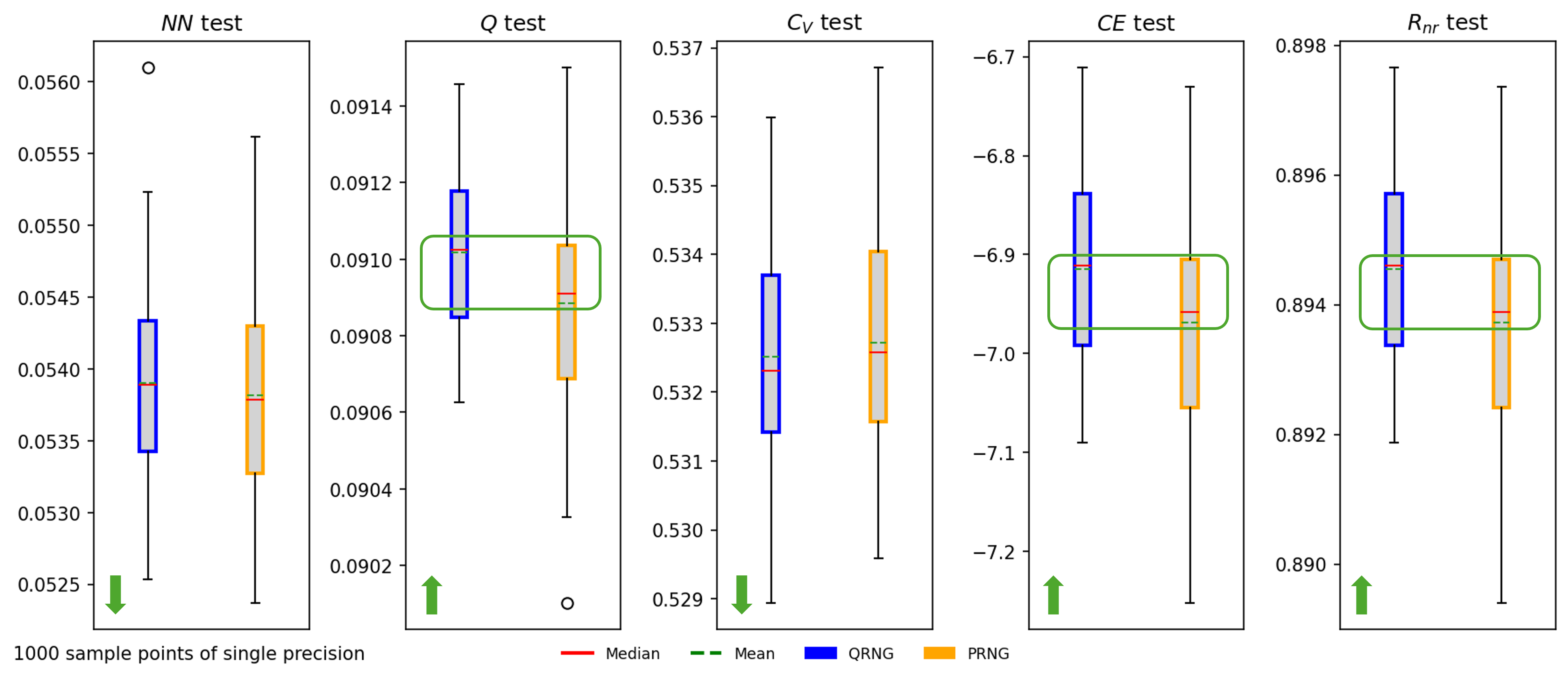}
    \caption{Uniformity measures for 1000 QRNG and PRNG float points.\\ The green arrow indicates the direction of more uniform sampling, the green box highlights cases that are statistically significant. Higher $Q$ and $R_{nr}$ with values $<1$  indicate a tendency of QRNG towards more randomness with less inhibition, as evidenced by $C\!E$ test.}
    \label{fig:1K_statistics}
\end{figure}

\begin{table}[ht]
\centering
\begin{tabular}{ccccc}
\hline
            & W test 							& 	T test 							&  AD test						& KS test \\ \hline
$N\!N$      & 0.5210 							& 	0.5286 							&  0.2500 						&  0.8693 \\
$Q$         & \underline{0.0171}	& 	\underline{0.0141} 	&  \underline{0.0155}	&  \underline{0.0217} \\
$C_V$       & 0.5148 							& 	0.5577 							&  0.2500 						&  0.9977 \\
$C\!E$      & \underline{0.0207} 	&  	\underline{0.0156} 	&  \underline{0.0183} &  \underline{0.0392} \\
$R_{nr}$    & \underline{0.0207} 	&  	\underline{0.0156}	&  \underline{0.0183} &  \underline{0.0392} \\
$LES$       & 0.6390 							&  	0.6575 							&  	0.2500 						&  0.5487\\ \hline
\end{tabular}
\caption{$p$-values for 1000 float points, $\alpha=0.05$}
\label{tab:1K_single}
\end{table}

In \ref{fig:1K_statistics} we present results for the 
 $50\times 100$ sets of 1000 points, each of which is defined by a single-precision number pair representing the $x$- and $y$-coordinates. Although the direct application of  $N\!N$ measure does not illustrate any statistically significant difference, the measures derived from  $N\!N$ lead to the following observations: the means and medians of the quality measure $Q$ and $R_{nr}$ are significantly higher for QRNG as compared to PRNG but remain below 1, suggesting that QRNG offer a better dispersion of values, indicating a tendency towards more randomness. The observation that the Coefficient of Variation ($C_V$) is smaller for QRNG than for PRNG, despite no statistically significant difference observed in $N\!N$ distances,  \ref{fig:1K_statistics}, indicates that while both types of generators produce uniformly distributed numbers, QRNG does so with consistency and less variability more closely matching theoretical expectations. The statistically significant differences are observed at a $95\%$ confidence level between the sets of single-precision floating-point numbers generated by QRNG and PRNG, as determined by the Wilcoxon, Student's T-test, Anderson-Darling, and Kolmogorov-Smirnov tests, and shown in \ref{tab:1K_single}.
 
There are fewer statistically significant observations for random number sets of medium size ($10K$ points of single or double precision, c.f. SM \ref{tab:10K_single}). In the case of the random double precision number sets of larger size ($50K$ points), there is a significant difference in the means of $LES$ radii between QRNG and PRNG, where the smaller radius for QRNG indicates that the points are more uniformly distributed without large gaps (c.f. SM \ref{tab:50K_single}, SM \ref{fig:50K_double_statistics}). The F-test results reveal that variance differences between QRNG and PRNG datasets are not statistically significant across the measurements, suggesting similar dispersion characteristics when larger number-of-points per unit square are used: explicitly, this resembles 'over-sampling' of the domain, where differences between sampling vanish.

We have demonstrated differences in the sampling itself, as well as the potential of QRNGs to yield performance advantages in terms of improved accuracy or reduced sample size in the examplary MC simulations. 
Higher accuracy, or reductions in sample size without loss of accuracy, should benefit every stochastic simulation (e.g., Black-Scholes, stochastic molecular dynamics), but we expect greatest benefits for problems where random numbers are essential, but do not dominate the overall computation time.

Additionally, given the observations of an increased homogeneity of samples, we anticipate that non-linear problems will enhance the observable differences between the random number generators. Further, we expect the differences to be emphasized in problems entailing estimation of rare event probabilities and general sampling of heavy-tailed distributions.


\section{Conclusions}
In this work, we assessed the effect of various entropy sources on the outcomes of stochastic simulations. Herein, we assembled a test suite based on Monte Carlo simulations, and a palette of statistical tests, with varying underlying assumptions, to compare a quantum random number generator (QRNG) to pseudo-random number generators (PRNGs) in serial and parallel processes. Stochastic estimation of $\pi$ was chosen due to its simplicity and availability of an exact solution. Using the simple $\pi/4$ estimate (method A) and testing the hypothesis that the median of the data is $\pi$, we observed a rejection after $\sim 1.14\cdot 10^6$ samples when utilising the MT PRNG. In contrast the QRNG passed the test even at $2.8\cdot 10^6$ samples. We thus conclude that a higher accuracy can be achieved when using the QRNG. 

Likewise, using Buffon's needle experiment, we demonstrated that the QRNG results pass the t-test (hypothesis: mean = $\pi$, data normally distributed) for sample sizes up to $4.54\times$ larger than those achieved by the parallel PRNG, thus resulting in a $\sim2\times$ better approximation of $\pi$. If the limit is relaxed to $1/\sqrt{N}$, QRNG permits up to $3.16\times$ larger samples, resulting in an up to $2.9\times$ better approximation. In the process, we have demonstrated the capability of this method to distinguish the effects of points-per-sample $M$ and sample number $N$, which is not commonly found in the literature. We have also shown that the use of a QRNG results in better approximations (lower approximation errors) for sub-optimal parameter choices ($N\gg M$), enabling large reductions ($\sim 8\times$) of the number of samples without loss of accuracy. Finally, we demonstrated that comparisons of the effects of QRNGs and PRNGs on MC simulations should be performed with industry-standard PRNGs (e.g. Mersenne Twister or recommended pPRNGs). 
Additionally, based on a uniformity analysis, we assessed differences in the random sampling underpinning the MC simulations: our findings suggest that the QRNG, especially at small sample sizes, offers a better dispersion of samples than the PRNG indicating a tendency of the QRNG towards more uniformly distributed samples, closely matching theoretical expectation. 

Moving beyond the investigated use cases with known solutions, we envisage, and encourage, further testing and applications of QRNGs in stochastic modelling, ranging from Bayesian inference, stochastic differential equations, optimisation, and Monte Carlo simulations to leverage the revealed advantages in all affected fields. We expect the differences between PRNGs and QRNGs to be more pronounced for non-linear problems, such as non-linear stochastic differential equations, and especially for workloads that rely upon probability estimates rather than the leading moments of a distribution. We plan to pursue a wider comparison of QRNGs, classical true random number generators (TRNGs) and a variety of PRNGs in the aforementioned fields using the methodology utilised in this paper.

\section{Acknowledgments}
We thank the tech team at Quantum Dice and, in particular, Ramy Shelbaya and Marko von der Leyen for stimulating discussions and input. This work was supported by Innovate UK under Project No. 10031836. 

\section{Competing financial interests}
\textbf{Quantum Dice Ltd.} The QRNG used in this work has been developed by Quantum Dice Ltd. \\
\textbf{HSBC}: The authors declare no conflict of interest. This paper was prepared for informative purposes and is not a product of HSBC Bank Plc. or its affiliates. Neither HSBC Bank Plc. nor any of its affiliates make any explicit or implied representation or warranty, and none of them accept any liability in connection with this paper, including, but limited to, the completeness, accuracy, and reliability of information contained herein and the potential legal, compliance, tax, or accounting effects thereof. This document is not intended as investment research or investment advice; or a recommendation, offer, or solicitation for the purchase or sale of any security, financial instrument, financial product, or service; or to be used in any way for evaluating the merits of participating in any transaction.
\section{Author contributions}
\textbf{Conceptualisation}: all authors. \textbf{Methodology}: statistical analysis: AL, OY; hardware: AM. \textbf{Software}: testbench: AL; interfacing: AM, AL. \textbf{Validation}: AL, OY. \textbf{Formal analysis}: AL, OY. \textbf{Investigation}: all authors. \textbf{Resources}: all authors. \textbf{Data curation}: AL. \textbf{Writing}: AL, AM, OY. \textbf{Project administration}: AM, AL, DR. \textbf{Funding acquisition}: all authors.

\section{Supplementary materials}
The Supplementary materials provide details about the methods used, additional investigations into the differences between the MC methods for estimating $\pi$, as well as additional results for the uniformity analysis using larger sets of points on the unit square.\\

\appendix
\section{Methods}
\subsection{Random number generators} 
The effect of different entropy sources are investigated based on the uniformity analysis and the outcomes of MC simulations. Specifically, a self-certifying quantum random number generator (QRNG) is compared against two pseudo-random number generators (PRNGs), namely the Mersenne Twister as the industry standard for sequential computing and Tina's random number generator for parallel processes.

\subsubsection{Pseudo-random number generators} \label{PRNGmethod}

A PRNG is an algorithm that, given some parameters, will yield a number from a periodic sequence of numbers. Generally the
parameters are a \textit{seed} and the iteration number, although the latter is often kept as an internal counter of the
RNG object. The seed determines the starting position within the sequence. A fixed seed will always result in the same sequence of numbers. Usage of a fixed seed is beneficial for reproducibility but highly questionable for the purposes of statistics.

The Mersenne Twister, also known as MT19937, is implemented in the standard C++ libraries and deemed as a common PRNG for stochastic simulations, mainly due to its long period of $2^{19937}-1$. We must remark that we deliberately avoid choosing the \verb|default_random_engine| in our implementation, as it is implementation defined\footnote{On the authors' system it defaults to a linear congruential engine with a $2^{31}-1$ period.}. The standard implementation of the MT is used for method A of the approximation of $\pi$. 

Method B is run in parallel processes within this work. Importantly, the Mersenne Twister is not recommended for use in parallel settings due to potential seeding issues, hence Tina's random number generator is deployed \cite{bauke2011tina_SI, Tina-Bauke_PhysRevE}. This PRNG is designed for parallel programming environments and equally suited for shared and distributed memory computers. We note explicitly that MT can be used in a parallel environment but its documentation \cite{StdMersenne_twister_engineCppreference, RandomGeneratePseudorandom_SI} does not warn the user against such.
With the only control parameter being the seed it is not obvious whether the resulting sub-sequences will be non-overlapping.

\subsubsection{Quantum random number generator}\label{QRNGmethod}

The employed quantum random number generator (QRNG) leverages the quantum mechanical behaviour of photons: when encountering a beam splitter, the photon state’s path is inherently unpredictable. One can thus measure the difference between the beam splitter outputs to generate random numbers. Additionally, with knowledge of the input photon state, or in other words, the number of photons, a certain level of entropy from the output differences can be guaranteed, allowing a source-device-independent protocol \cite{PhysRevX.10.041048_SI}. The setup implementing this protocol consists of an untrusted photon source (a laser), two beam splitters, and three detectors. The first beam splitter partitions a portion of the incoming light to certify the photon number at any given time. The remaining output from the first beam splitter passes through the second beam splitter. Two detectors are placed at the outputs of the second beam splitter, and the differencing signal between them is postprocessed to generate random numbers, with the minimum entropy guaranteed for every sample. Importantly, this setup allows for eliminating any classical noise when the random bits are generated: this process is known as self-certification and is carried out in real time in the device. This allows for a reliable assessment between the QRNG and PRNGs, as the random numbers stem from the quantum mechanical behaviour of photons instead of any potential residual classical noise. \\
The output of the QRNG is a stream of random bits which is transformed into a list of values. The implementation to call for random numbers directly uses the C++ standard library \cite{CPPLibr_SI}. The list of values conform to the distributions of random numbers specified in the C++ programming language standard, and random numbers can therefore be sampled from all distributions standardised in the C++ library.

\subsubsection{On Reproducibility}\label{subsec:Reproducibility}
To facilitate the collection of data required to apply the tests presented in \ref{subsec:Tests}
to each model, we employ Snakemake, \cite{molder2021}.
The workflows for the $\pi$-example are set up in a way to allow reproducible runs with (p)PRNGs
as well as QRNG, provided the entropy of the latter is stored in binary format to file.

It should be noted that the collection of one set of $\pi$ samples requires roughly $12.8$ TiB of
entropy. This effectively prevents a bit-wise fully reproducible set-up.

\subsection{Number $\pi$ approximation}\label{subsect:pi_approximation}
This section focuses on a single test case and two Monte Carlo methods to further investigate the quality of random numbers. We consider the simple method (Method A) and Buffon's needle technique (Method B) to find the approximation of $\frac{\pi}{4}$ and $\frac{2}{\pi}$, respectively. Method A has been considered as a part of an emerging test suite \cite{almarazluengoStringENTTestSuite2023},
whereas method B is generally relegated only to academic examples. Below we present the mathematical formulation of the methods and the error estimates for the approximations of number $\pi$. Via the upper subscript, we indicate the reference to the corresponding method.

Let an indicator random variable $X_i^A$ represent the outcome of whether a randomly chosen point $(a,b)$ falls inside the first quarter of the unit disk, $a,b \sim U[0,1]$
\[X_{i}^A = \begin{cases}
1, & \sqrt{a^2+b^2}\leq 1, \\
0, & \text{otherwise}
\end{cases}, \quad p^A=P(X_{i}^A) = \frac{\pi}{4} \]
or let $X_i^B$ represent the outcome of Buffon's needle experiment.
The needle is characterized by the distance from the center to the nearest line, $\rho \sim U[0,1/2]$, and the angle $\theta \sim  U[0,\pi/2)$  relative to the line.
\[ 
X_{i}^B = \begin{cases}
1, & \rho<1/2\sin\theta,\\
0, & \text{otherwise } 
\end{cases}, \quad p^B=P(X_{i}^B) = \frac{4}{\pi}\int_0^{\pi/2} d\theta \int_0^{\frac{1}{2}\sin\theta} d\rho= \frac{2}{\pi}. 
\]
If we denote by $p$ the probability of the event $X_i$, then  $X_1, X_2, \ldots, X_M$ is the sequence of independent and identically distributed random variables with finite mean ${\mathbb E}[X_i] = p$ and finite variance $\text{Var}(X_i) = p(1-p)$. 

To derive the approximation for the number $\pi$ and the corresponding error estimate we define the random variable $Y_M = X_1+\ldots + X_M$ which represents the total number of positive outcomes out of $M$ trials. Due to the law of large numbers applied to $Y_M$, as \( M \) increases, the average of \( Y_M \) converges to the expected value, which can be related to the estimate of $p$ via \[\hat p = \frac{k}{M},\] 
where \( k \) represents the successes in \( Y_M \). Hence, the approximations to $\pi$ are given by
\[\hat{\pi}_{A} =  4\hat{p}, \quad \hat{\pi}_{B} = 2/ \hat{p}.\]
Recalling \[{\mathbb E}[Y_M] = M{\mathbb E}[X_i]=Mp, \quad \text{Var}(Y_M)=M\text{Var}(X_i)=Mp(1-p), \]
the central limit theorem yields $\displaystyle Z = \frac{k - p M}{\sqrt{M p(1-p)}} \sim \mathcal{N}(0, 1)$  for sufficiently large $M$, and we find the upper bound
\begin{equation}\label{estimate:1}|\hat{p} - p| \leq \frac{\Phi^{-1}(1 - \frac{\alpha}{2})}{\sqrt{M}} \sqrt{p(1-p)},\end{equation}
where $\Phi^{-1}$ is the inverse of the cumulative distribution function (CDF) associated with the standard normal distribution, $Z \in \left[ -\Phi^{-1}\left(1 - \frac{\alpha}{2}\right), \Phi^{-1}\left(1 - \frac{\alpha}{2}\right) \right]$ with probability $1-\alpha$.
Finally, from (\ref{estimate:1}) we obtain the error estimate for the number $\pi$ approximation
\begin{equation} \label{eq:estimate_2}
|{\hat\pi}_A-\pi| \leq \pi \sqrt{{4}/{\pi} - 1}\,\, \frac{\Phi^{-1}(1 - \frac{\alpha}{2})}{\sqrt{M}}, \; |{\hat\pi}_B-\pi| \leq \pi \sqrt{{\pi}/{2} -1}\,\, \frac{\Phi^{-1}(1 - \frac{\alpha}{2})}{\sqrt{M}}.
\end{equation}

We note that the accuracy of $\pi$ approximation might be limited by the number of sampling points $M$, i.e. a systematic (numerical) error $\epsilon_M$ occurs. 
\begin{figure}[ht!]
    \centering    \includegraphics[width=0.7\linewidth]{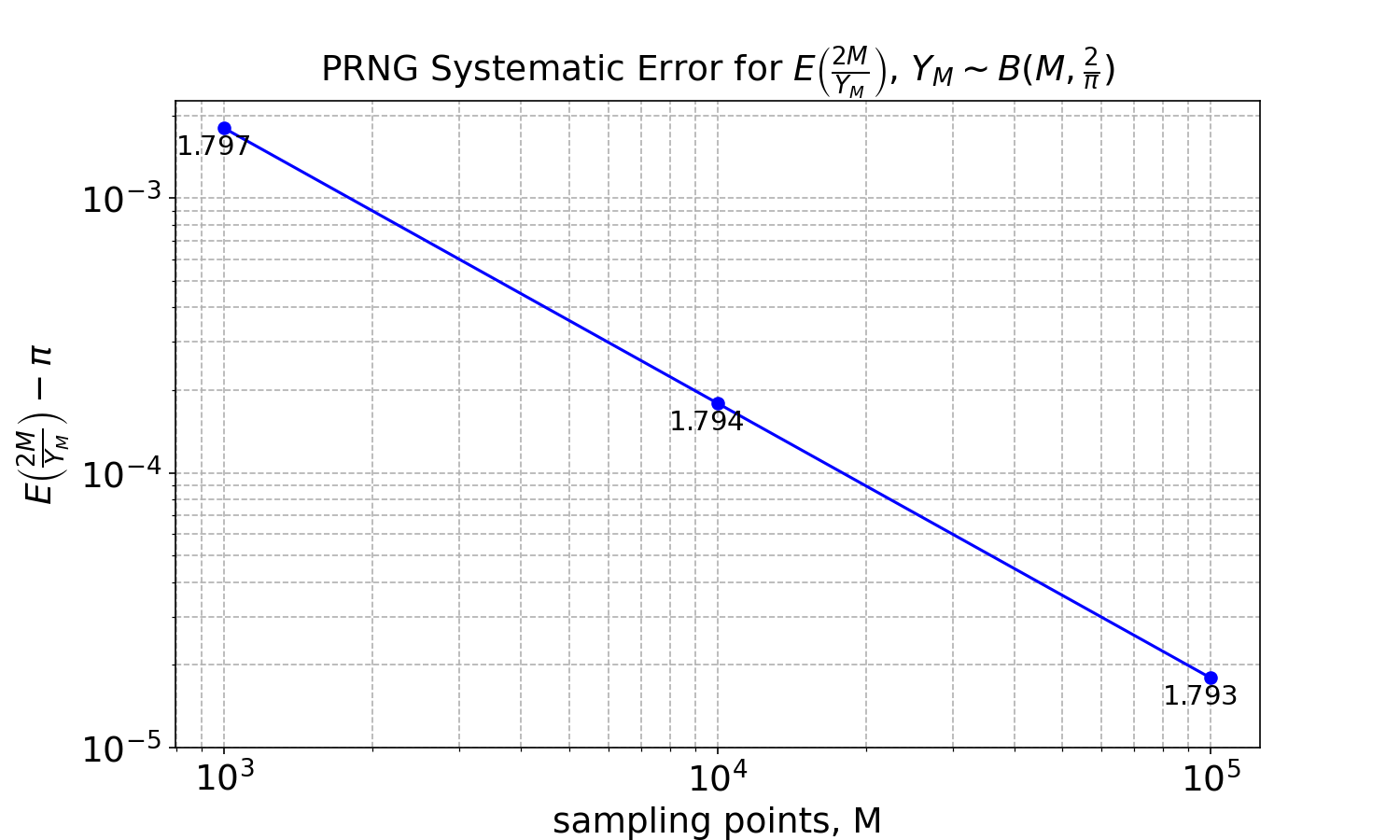}\caption{Systematic error in Buffon's needle experiment}
    \label{fig:systematic_error}
\end{figure}
 Indeed, considering the random variable $Y_M$ based on Buffon's needle method, one observes that $Y_M$ follows a binomial distribution skewed to the left with a skewness value of $\frac{({\mathbb E}[Y_M-{\mathbb E}[Y_M]])^3}{(\text{Var}[Y_M])^{3/2}}=\frac{1-2p}{\sqrt{M}\sqrt{p(1-p)}}\approx\frac{-0.5681}{\sqrt{M}}$. The impact of excluding the zero outcome on the distribution's skewness is minimal for large $M$. The negative skewness yields occurrences of smaller values of  $k$ which leads to $\pi<\overline{\hat{\pi}_B}$, and aligns with the results of Jensen's inequality
\[\frac{1}{{\mathbb E}\left[\frac{Y_M}{2M}\mid Y_M>0\right]}< {\mathbb E}\left[\frac{2M}{Y_M}\mid Y_M>0 \right]=\sum_{k=1}^{M}\frac{2M}{k}\,\binom{M}{k} \frac{p^k (1 - p)^{M - k}}{1-(1-p)^M}\]
resulting in 
\[\pi<\lim_{M\to\infty}{\mathbb E}\left[\frac{2M}{Y_M}\mid Y_M>0\right].\]
Although it is difficult to find the theoretical formula for ${\mathbb E}[\frac{2M}{Y_M}\mid Y_M>0]$ in a closed form, the presence of the systematic error, i.e. $\overline{\hat{\pi}_B}\approx \pi + \epsilon_M$, can be observed from the numerical values of Jensen's gap on  \ref{fig:systematic_error}, $\epsilon_M\approx 1.8/M$.

\subsection{Uniformity analysis}\label{subsec:uniformitymethod}

In this section, we provide metrics to quantify the spatial uniformity of two-dimensional data sets. Testing uniformity in the unit interval has been extensively studied and the overview of some of the methods may be found in \cite{marhuenda2005comparison}.
On the other hand, the uniformity in the multivariate setting is far less studied.
For the overview of statistical methods for analysing data in the form of spatial point distributions, we refer to \cite{diggle2013statistical}.

To assess whether a set of points in space follows a uniform distribution, one can examine the distances between pairs of points. Some authors consider the asymptotic properties of small inter-point distances or the distance to the nearest neighbor ($N\!N$), while others investigate the moments and distribution of the inter-point distances, \cite{ebner2018multivariate}.

We concentrate on the two metrics for the uniformity test of the set of $M$ points $\mathcal X_M=\{x_1,\ldots, x_M\}\subset [0, 1]^2$, $M\in \mathbb{N}$: the distance to the nearest neighbor (NN) 
\begin{equation}\label{NN}
\gamma_{max} = \max\limits_{i=1,...,M}\gamma_i, \quad \gamma_i = \min\limits_{j=1,...,M; j\neq i}|x_i-x_j|,
\end{equation}
and the radius of the largest empty sphere (LES) within the unit square, \cite{ebner2018multivariate},
\begin{equation}\label{LES}
\Delta_M = \sup\{r > 0 : \exists x\in [0, 1]^2 \mbox{ such that } B(x,r) \subset [0, 1]^2\setminus{\mathcal X_M}\}.
\end{equation}
Should there be deviations from uniformity via “clustered” observations, the maximal spacing or Janson’s uniformity test, \cite{berrendero2012multivariate}, on a unit square, would reject, at a significance level $\alpha$, the null hypothesis "random variable has a uniform distribution" whenever
\begin{equation}\label{MS_LES} \pi\Delta_M^2> (-\ln(-\ln(1-\alpha)) + \ln M + \ln \ln M)/M.\end{equation}

\begin{figure}[ht!]
    \centering
    \includegraphics[width=0.4\linewidth]{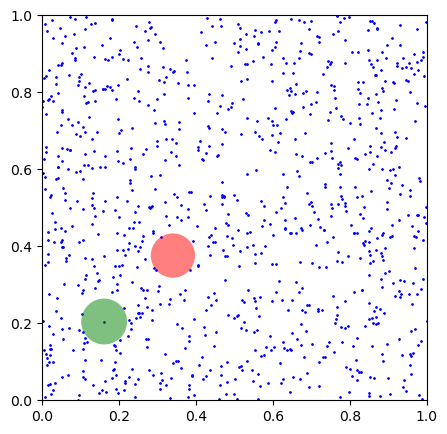}
    \hspace{0.5cm}
    \includegraphics[width=0.4\linewidth]{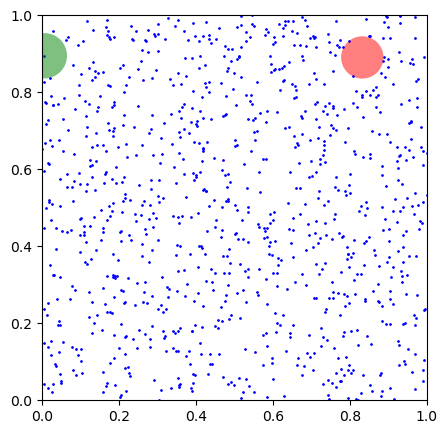}
    \caption{QRNG (on the left) and PRNG (on the right), 1000 float points.\\
{\hspace*{1.7cm} The green circles represent $N\!N$ and the red ones demonstrate the LES.}}
    \label{fig:uniformity_of_sampling}
\end{figure}

The metrics (\ref{NN}) and (\ref{LES}) are illustrated in  \ref{fig:uniformity_of_sampling} with the sampling of a unit square by QRNG and PRNG, on the left and right sides respectively. The green circles exemplify the distance to the nearest neighbour ($N\!N$) and the red ones present the largest empty sphere (LES).
Utilising the aforementioned metrics, we perform several statistical tests such as the Wilcoxon test and F-test and present the comparison results in \ref{sec:numerics}. 

Based on the $N\!N$  metric, we also apply the normally distributed test statistic $C\!E$ and the "measure of non-randomness" $R_{nr}$ suggested by Clark and Evans in 1954 and improved by Donelly in 1978, \cite{ripley1979tests}
\begin{equation}\label{CE_R}
C\!E = \frac{\bar{\gamma} - E(\bar{\gamma})}{\sqrt{\text{Var}(\bar\gamma)}}, \quad R_{nr} = \frac{\bar{\gamma}}{{\mathbb E}(\bar\gamma)},
\end{equation}
with $\bar\gamma=\displaystyle\frac{1}{M}\sum_{i=1}^M \gamma_i$, ${\mathbb E}(\bar{\gamma}) = 0.5{M}^{-1/2} +4M^{-1}(0.514 +0.412{M}^{-1/2})$ and $\text{Var}(\bar\gamma) = 0.07M^{-2} +0.148 M^{-5/2}.$
With these approximations for the mean value and the variance, the $C\!E$ is assumed to follow a standard normal distribution. Clark and Evans tests are powerful both in detecting clustered patterns and deviations from randomness in cases where the events are more evenly distributed across the spatial domain, i.e. a large positive $C\!E$ suggests clustering, while a small negative $C\!E$ indicates inhibition. \cite{mugglestone2001spectral}

Furthermore, we consider the coefficient of variation $C_V$ and the mesh ratio $R_M$ measures due to Gunzburger and Burkardt, 2004 and the potential energy quality measure $Q$ introduced by Ong et al 2012, \cite{ong2012statistical}:
\begin{equation}\label{CoV_MR}
C_V \!=\! \frac{1}{\bar\gamma} \sqrt{ \frac{1}{M} \sum_{i=1}^{M} (\gamma_i - \bar{\gamma})^2}, 
\, R_m \!=\! \frac{\max\limits_{i=1,...,M}\gamma_i}{\min\limits_{i=1,\ldots,M} \gamma_i}, 
\, Q\!=\!\frac{1}{M}\sum_{i=1}^M\left(\!1-\frac{{3}/{M}}{{3}/{M}+\gamma_i^2}\!\right),
\end{equation}
where the second term under the sum in $Q$ represents the potential energy between the $i$-th point and its nearest neighbor. The smaller the values $C_V$ and $R_m$ are, the closer the points are to the perfectly uniform mesh.
A higher $Q$ value suggests that points are more uniformly dispersed. Conversely, lower $Q$ values indicate that points are more clustered or unevenly distributed, with more instances of points being closer to their neighbours. By averaging the contributions from each point, $Q$ provides a metric that reflects the overall uniformity of the distribution.

The mesh ratio $R_m$ assesses the local uniformity of point sets, as noted in \cite{ong2012statistical}, whereas the coefficient of variation measure $C_V$ and the quality measure $Q$ provide a more global measure.

\subsection{Statistical tests}\label{subsec:Tests}
To determine whether any statistically significant differences between the samples obtained
with QRNG and PRNGs are present, we employ a palette of statistical tests \cite{sircaProbabilityPhysicists2016}[Ch. 10].
Herein, we retain tests with a variety of assumptions on the underlying distribution to investigate the difference between the distributions from multiple points of view.
For the $\pi$ model each dataset is tested individually against the known
result using a 1-sample test check whether the result is acceptable as an approximation of $\pi$. When no reference value is available, as is the case with the uniformity measures of \ref{subsec:uniformitymethod}, a 2-sample version of the tests is performed to determine whether the datasets could originate from the same distribution.

\paragraph{The sign test} tests the hypothesis that the median $m$ of the sample is equal to a given value $\mu_0$ - formally $H_0: m=\mu_0$ - against the alternative hypothesis $H_A:m\neq\mu_0$.
As such, its sole requirement is that the median can be computed, i.e., that the sample can be ordered.
A two-sample sign test checks the hypothesis that both samples possess the same median. It implicitly \textit{assumes} that the sample values ($x_i,y_i$) are paired s.t. a derived variable $u_i=x_i - y_i$ can be computed, only requiring the same number of samples. The $u_i$ are thus assumed to be samples of a random variable $U$ with a $0$ median.

Since the median partitions a dataset into 2 equal sizes, the number of points above/below the median should be equal.
This number, denoted here as $C$ or test statistic, follows a Binomial distribution with the number of samples being the number of trials and the success probability
$p=\tfrac{1}{2}$, i.e. $C\sim\text{Bin}\left( N,\frac{1}{2}\right)$. It is then trivial to determine the acceptance range of the number of values above/below the median to any significance level $\alpha$. In general, we employ a two-sided test.
The \textbf{Wilcoxon test} is an improvement upon the simple sign test but \textit{assumes} that the samples originate
from a symmetric distribution.

\paragraph{T-test,} or Student's t-test \cite{sircaProbabilityPhysicists2016}[Ch. 10.2.1], for a sample of $N$ values $x_i$ uses the test statistic
\begin{equation}
T := \frac{\bar{X}_N - \mu }{S_N / \sqrt{N}}\;,\label{eq:T_test_statistic}
\end{equation}
with $S_N$ being the empirical standard deviation to test whether the empirical mean $\bar{X}_N$ is statistically identical to the prescribed mean $\mu$. It is known that
the random variable $T$ will be distributed according to the t-distribution with $N-1$ degrees of freedom,
which allows us to determine the acceptance range for $T$ to a given significance level $\alpha$.
The statistic is derived from the so-called $z$-test and hence fundamentally assumes a normal distribution of the samples $x_i$.

\paragraph{F-test,} also known as the variance ratio test, is a statistical method used to assess the equality of their population variances. The F-test assumes that the data follow a normal distribution, are independent, have equal variances across groups, and are measured on an interval or ratio scale.
The test is performed on the quotient
\begin{equation}
F := \frac{ s^2_{1}}{s^2_{2} }\;,\label{eq:F_test_statistic}
\end{equation}
where  $s^2_{1}$, $s^2_{2}$ are the variances of the samples. If both variances are computed using the same number of samples $N$, the F-statistic follows the F-distribution characterised by 
$N-1$ degrees of freedom for both the numerator and the denominator \cite{sircaProbabilityPhysicists2016}[Ch. 10.2].

\paragraph{CDF-based tests,} such as the "Kolmogorov-Smirnov" (KS) and "Anderson-Darling" (AD) 2-sample tests compare the cumulative distribution functions (CDF) for both datasets. Both order the data $x_i$, construct an empiric cdf $F_{1,2}(x)$ based on said data and yield $\sup_{x\in\mathbb{R}} \Vert F_1(x) - F_2(x)\Vert$. The closer the two distributions are, the smaller the statistic will be and an acceptance threshold $d^{\ast}(N,\alpha)$ can be determined numerically for a sample size $N$ and confidence parameter $\alpha$. The AD test, instead, determines the weighted $L^2$ distance between the distributions, with the weighting function $\omega$ chosen to emphasise the tails of the distribution.

\section{Numerical Experiments}\label{sec:numerics}

To analyse the effects of the random number source, we have performed a wide range of numerical experiments. Each of the uniformity experiments and $pi$ approximations was performed using numbers in single-(\verb|float|) and \verb|double|-precision, to account for any potential differences in grouping of random bits.
The $\pi$ data is collected by supplying the simulation routines the number of points per sample $M$ as well as the number of samples per batch $N_{spb}$ and the total number of batches $N_{batches}$. Batching is used for a mean-of-means calculation during the simulation, and has no effect on the stored samples. For the total number of samples we use $N=N_{spb}\cdot N_{batches}$.
Batch size variations serve to distinguish individual executions of the program in \ref{fig:EmpiricMeansEvolution}, each of which corresponds to a PRNG initialisation with a different seed.

If no seed is prescribed for a (p)PRNG, the current time at the start of the execution is used. With this approach we directly contradict textbook suggestions \cite{grahamStochasticSimulationMonte2013}(Ch. 2.2.1) of using a fixed seed. This decision was made to represent a general case, avoiding potentially biasing the findings by choice of a seed that results in a particularly "good" subsequence (e.g., in particularly small variance of the data), as well as to match the non-deterministic behaviour of the QRNG. 

To further reflect the application of MC simulations in practice, our simulations are performed using 4 OpenMP threads. In the case of the MT PRNG, the seed of each thread is offset from the common seed by the thread index, and for the pPRNG, the leapfrog drawing method is used\cite{bauke2011tina_SI}.
In exploring the effects of the RNG, we have considered various points per sample (pps) $M$, sample sizes $N$ and precisions of the fundamental data (\verb|float,double|).

\subsection{On the number $\pi$ approximations}\label{sec:Buffonfinding}

Since Buffon's method approximates $\frac{2}{\pi}$ instead of $\frac{\pi}{4}$, the behaviour of the $\tilde{\pi}$ distributions is different.
\begin{figure}[htb]
    \centering
    \includesvg[width=\linewidth]{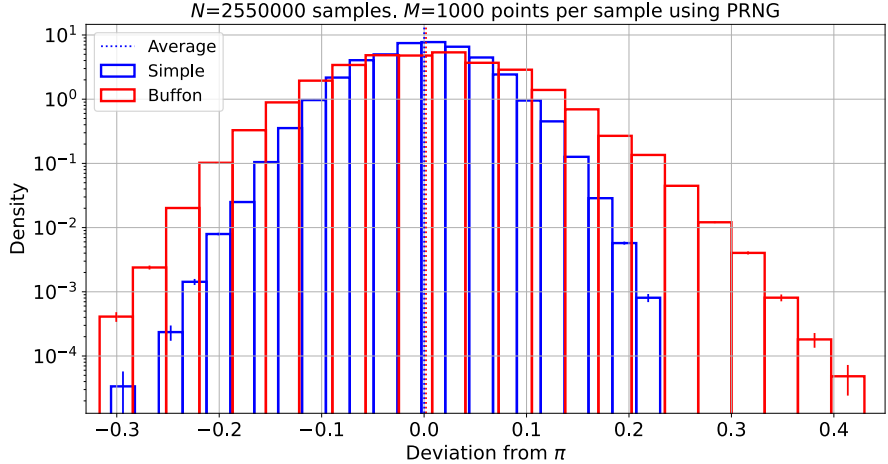}
    \caption{Distribution of $\tilde{\pi} - \pi$ differences for samples obtained with methods A, B using $N=2.55\cdot 10^6$ samples and $M=10^3$ pps.}
    \label{fig:SimpleVsBuffonDistribution}
\end{figure}
In \ref{fig:SimpleVsBuffonDistribution} we show the event density histogram of $\tilde{\pi} - \pi$ for methods A (blue)  and B (red) using $1000$ points per $\tilde{\pi}$ sample, both using a pPRNG. It is immediately obvious that the variance of the simple approximation is smaller. Upon closer inspection, one can see that the mean obtained with Buffon's experiment differs from the true value of $\pi$ more than the mean obtained with method A. This indicates that method A is the better choice for approximating $\pi$ via MC simulations.

\begin{figure}[htb]
    \centering
    \includegraphics[width=0.6\linewidth]{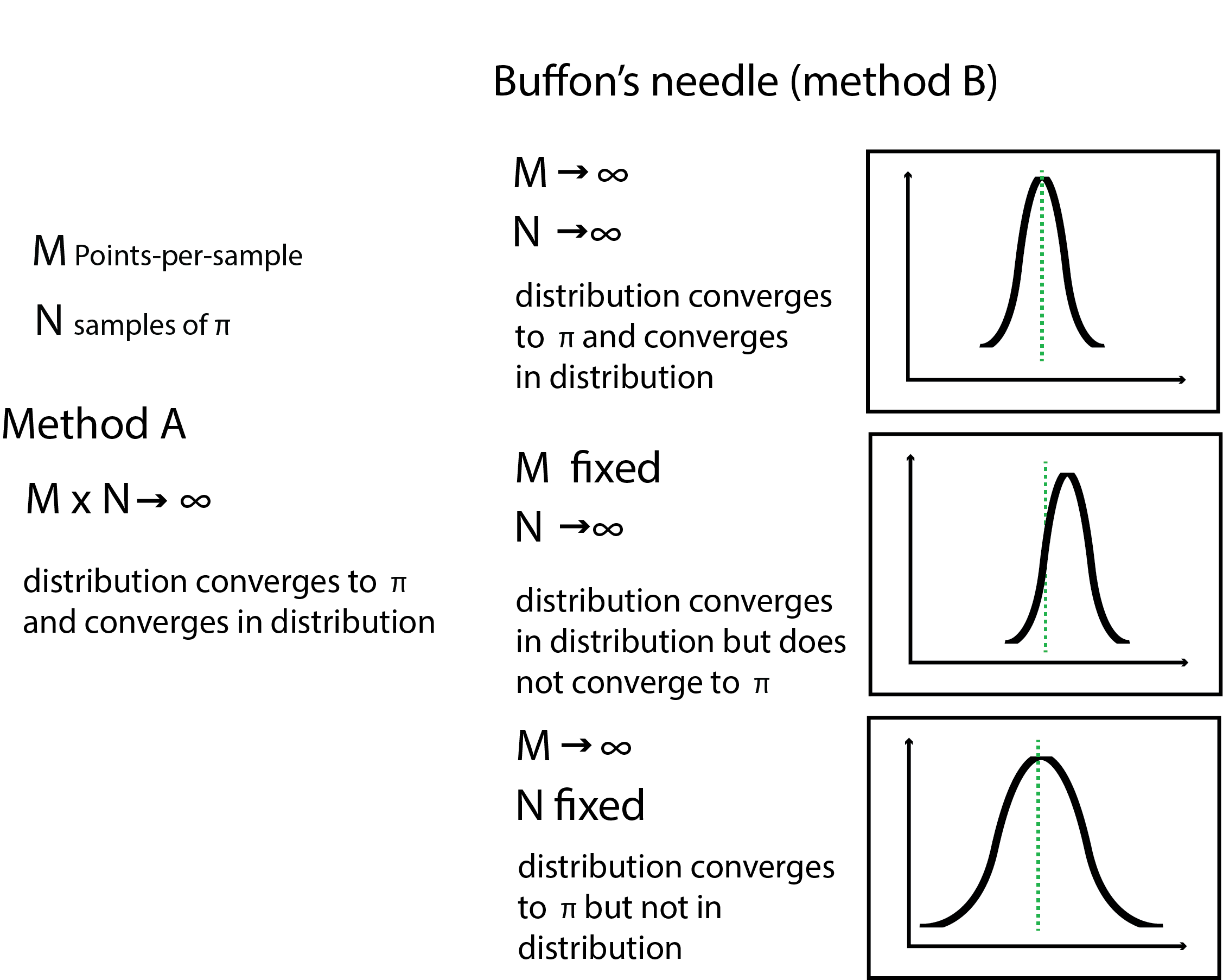}
    \caption{Convergence behaviour w.r.t $M$ points-per sample and $N$ samples.}
    \label{fig:Convergencebehaviour}
\end{figure}
However, the approximation error of method A displays no clear dependence on $M$ and is hence assumed independent of $M$.
In contrast, Buffon's method yields larger approximation errors for $M\leq 5\cdot10^5$ but displays the expected $\frac{1}{\sqrt{M}}$ convergence, resulting in a better approximation at $M=10^6$ pps and beyond. Hence, method B allows for a clear distinction of the effects of the sample size $N$ and the points per sample $M$, whereas for method A such a distinction is not easily made (\ref{fig:Convergencebehaviour}).
\subsection{Uniformity analysis - Additional results}
The measures described in \ref{subsec:uniformitymethod} are collected 50 times for the sample points sets of various sizes and numerical precision both for sequential pre-generated QRNG and PRNG (Mersenne Twister).

\begin{table}[ht]
\centering
\begin{tabular}{cccccc}
\hline
            & W test 							&  T test 							&  F test 						&  AD test							& KS test \\ \hline
$N\!N$      & 0.3522 							&  0.2606 							&  0.7260							&  0.2500 							&  0.7166 \\
$Q$         & 0.1305 							&  0.1007 							&  0.3009 						&  0.0648 							&  0.1124 \\
$C_V$       & 0.1719 							&  0.1652 							&  \underline{0.0165} &  0.0548 							&  \underline{0.0392} \\
$C\!E$      & {0.0789} 						&  {0.0688} 						&  0.1508 						&  \underline{0.0272} 	&  \underline{0.0392} \\
$LES$       & 0.7377 							&  0.6263 							&  0.7106 						&  0.2500 							&  0.9667\\ \hline
\end{tabular}
\caption{$p$-values for 10000 double points, $\alpha=0.05$}
\label{tab:10K_single}
\end{table}

\begin{figure}[ht!]
    \centering
    \includegraphics[width=1\linewidth]{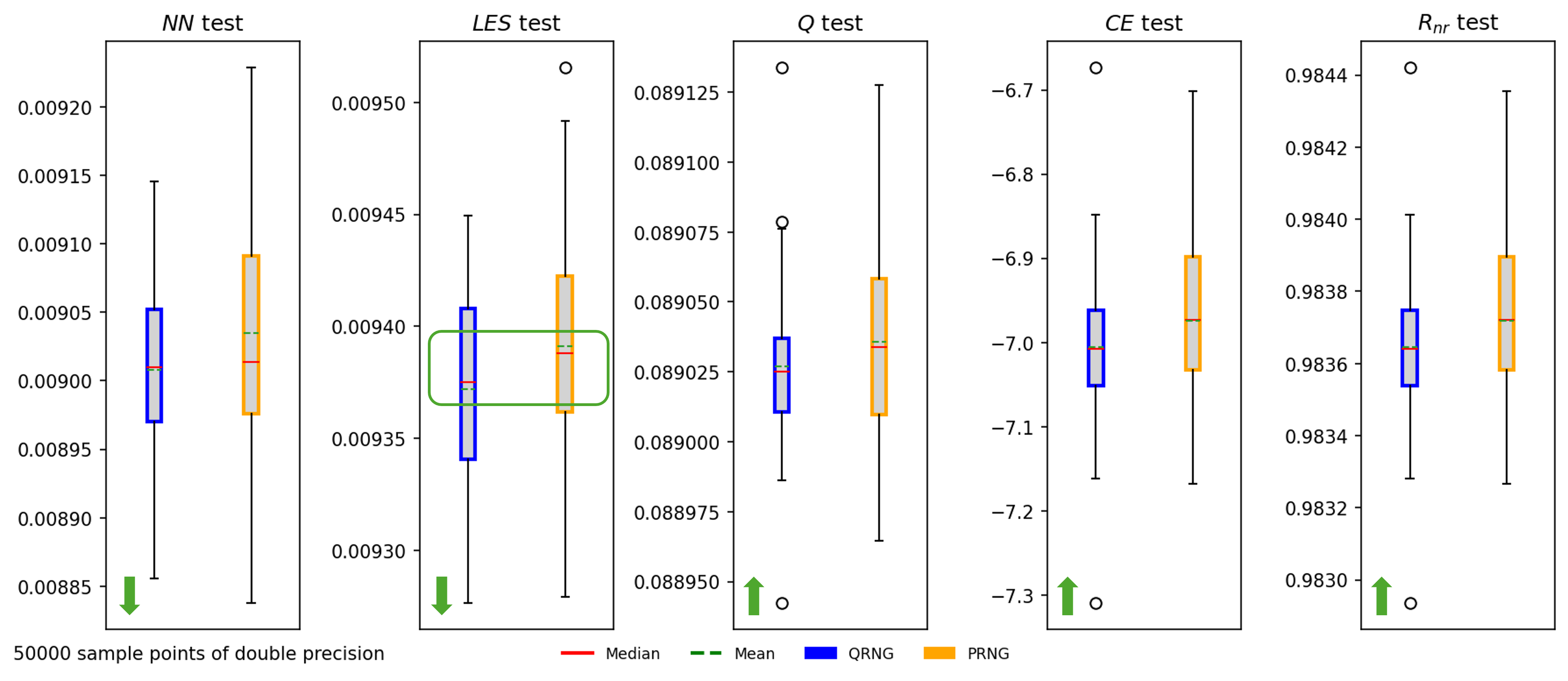}
    \caption{Uniformity measures for 50000 QRNG and PRNG double points.\\
    The green arrow indicates the direction of more uniform sampling, the green box highlights cases that are statistically significant. Lower $LES$ indicates that QRNG points are more uniformly distributed without large gaps.}
    \label{fig:50K_double_statistics}
\end{figure}

\begin{table}[ht]
\centering
\begin{tabular}{cccccc}
\hline
            & W test 						&  T test 							&  F test 						&  AD test							& KS test \\ \hline
$N\!N$      & 0.2335 						&  0.0842 							&  0.9157 						&  0.1949 							&  0.2719 \\
$Q$         & 0.1844 						&  0.1863 							&  0.8941 						&  0.0722 							&  0.0678 \\
$C_V$       & 0.3137 						&  0.2805 							&  0.1935 						&  0.2500 							&  0.5487 \\
$C\!E$      & 0.1139 						&  0.1149 							&  0.7759 						&  0.0816 							&  0.0678 \\
$LES$       & 0.0518 						&  \underline{0.0469}		&  0.7863 						&  0.1008 							&  0.2719 \\ \hline

\end{tabular}
\caption{$p$-values for 50000 double points, $\alpha=0.05$}
\label{tab:50K_single}
\end{table}



\clearpage
\end{document}